# Journal Name

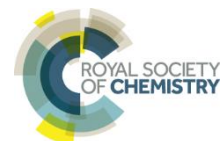

## COMMUNICATION

# Mechanoluminescent MOF nanoplates: spatial molecular isolation of light-emitting guests in a sodalite framework structure

Abhijeet K. Chaudhari and Jin-Chong Tan*

Mechanoluminescent materials have a wide range of promising technological applications, such as photonics based sensors and smart optoelectronics. Examples of mechanoluminescent metal-organic framework (MOF) materials, however, are relatively uncommon in the literature. Herein, we present a previously undescribed Guest@MOF system, comprising "Perylene@ZIF-8" nanoplates which will undergo a reversible 442 nm ⇌ 502 nm photoemission switching when subject to a moderate level of mechanically induced pressure (~10s MPa). The nanoplates were constructed via high-concentration reaction (HCR) strategy at ambient conditions, to yield a crystalline ZIF-8 framework hosting the luminous Perylene guests; the latter confined within the porous sodalite cages of ZIF-8. Remarkably, we show that in a solid-state condition, it is the spatial isolation and nano-partitioning of the luminescent guests that bestow a unique solution-like optical properties measured in the host-guest assembly. As such, we demonstrate that switchable red- or blue-shifts of the visible emission can be accomplished by mechanically modifying the nanoscale packing of the nanoplates (e.g. monoliths, pellets). Theoretical calculations suggest that the elasticity of the host's sodalite cage coupled with the intermolecular weak interactions of the confined guest are responsible for the unique mechanoluminescent behavior observed.

## Introduction

Functional nanomaterials with adjustable luminescent properties[1-3] are greatly sought after to achieve new generation photonics-based sensing and tunable optoelectronics applications.[4-6] Indeed, a vast range of fluorescent materials, including conjugated organic molecules, metal complexes, liquid crystals, oligomers, polymers, metallogels, supramolecular assemblies and quantum dots have shown huge potential for use in many luminescent based devices and sensors.[7-10] One of the most intriguing findings in this topic area is *mechanoluminescence*, reported way back in 1605 by Francis Bacon, who observed that lumps of sugar emitting light when crushed.[11] Mechanoluminescent materials emit light when subject to mechanical forces (stress).[12] They have gained increasing attention in recent years, where researchers seek to understand, control and tune luminescent properties with the help of exogenous mechanical stimuli, ranging from shearing, grinding and rubbing, to tension, compression and impact. It is envisaged that smart mechanoluminescent materials can be integrated into deformation-based detectors, security papers, optical memories, and photonic strain gauges.[13, 14] Regarding material design, it is essential to control its intermolecular weak interactions, which are dictating the optical behavior as a function of mechanical stress and strain.[15, 16] Outstanding challenges in the field include aggregation caused quenching (ACQ), solid-state phase change and amorphisation, limited structural resilience and difficulty in the precise control of molecular orientations needed for practical applications. Traditional approaches for making mechanoluminescent materials often require complicated molecular designs and will involve difficult multistep synthetic procedures.[17-20] To address the foregoing challenges, one possible solution is to combine chemical traits of existing fluorophoric molecules into ordered nanoscale host assemblies, by leveraging the host-guest spatial confinement strategy.[21-24]

Metal-organic frameworks (MOFs)[25, 26] have been intensely studied for over two decades for their high porosity and long-range crystalline network, combined with designable chemical environments and tunable physical properties.[27-30] To trigger switchable optical properties of guest molecules under the confinement of the nanoscale pores in MOF host, dynamic MOF systems is highly desirable owing to their structural versatility. Recently we shed light on the dynamics of zeolitic imidazolate frameworks (ZIFs: is a topical class of MOFs)[31] via Terahertz vibrational spectroscopy,[32] and demonstrated the efficacy of the sodalite cage of ZIF-8 for hosting bulky metal complexes to engineer luminescent material with improved photostability.[23] In this study, we combine our understanding of the dynamic behavior of ZIF-8 framework and its use as a host for accommodating external guest species to contrive a new mechanoluminescent host-guest material. There are already a few examples of (host-only) mechanoluminescent MOFs reported in the literature (e.g. refs.[33-37]) in which mechanoluminescence originated typically from the N-donor

*a*Multifunctional Materials & Composites (MMC) Laboratory, Department of Engineering Science, University of Oxford, OX1 3PJ, Oxford, United Kingdom. Email: jin-chong.tan@eng.ox.ac.uk

†Electronic Supplementary Information (ESI) available: [Detailed synthesis procedures; microstructural characterisation by TEM and AFM; TGA data; photoluminescence spectra of pellets pressed at different pressures; band gap calculations; XRD structural refinement results.]. See DOI: 10.1039/x0xx00000x

based coordinating linkers. To the best of our knowledge, the host-guest system reported herein is the first example of a mechanoluminescent material demonstrating spatially ordered fluorophoric guests constrained in a periodic MOF structure.

## Results and discussions

### 1. Guest@MOF Encapsulation Achieved by High Concentration Reaction

Facile reaction of 2-methylimidazole (mIm) with Zn(II) cations yields the formation of ZIF-8 adopting the sodalite framework topology.[38] ZIF-8 is a highly topical MOF structure for encapsulation of functional guest molecules, for example see refs.[23, 39-46], because it features long-range periodicity with a relatively large pore (~1 nm diameter) located in the middle of each sodalite cage. The ease of synthesis combined with the unique pore architecture and mechanical properties of ZIF-8 (elasticity[47] and dynamics),[32] could offer vast possibilities to engineer novel Guest@MOF systems exhibiting tunable functionalities.

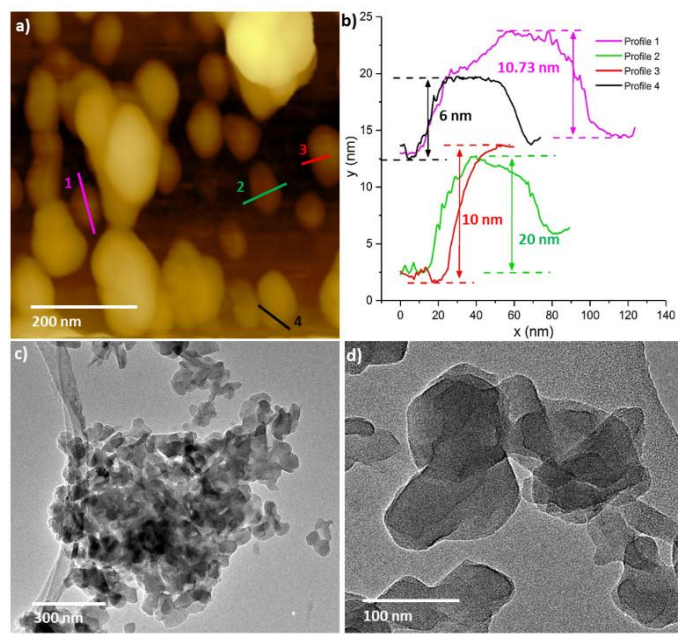

**Fig. 1** (a) AFM topography image of nanoplates of Perylene@ZIF-8. (b) Representative thickness profiles of nanoplates obtained from the AFM image in (a), indicating a typical thickness of the order of ~10s nm. (c) TEM micrograph showing the aggregated nature of the finer scale Perylene@ZIF-8 nanoplates. (d) A higher-magnification TEM image showing the intertwined 2-D microstructural features of the nanoplates.

In this study, we have exploited the spherical voids of ZIF-8 as spatially arranged "host" spaces, allowing encapsulation of the polycyclic aromatic hydrocarbon: Perylene ($C_{20}H_{12}$), which is an efficient light-emitting "guest" molecule. We adopted the concept of high-concentration reaction (HCR) introduced recently by our group,[22, 48] to synthesize a new host-guest assembly termed: Perylene@ZIF-8, at ambient conditions. The detailed synthetic procedures are described in the Supporting Information (SI). In essence, the proposed self-assembly route enables us to achieve nanoscale spatial confinement of bulky Perylene guest molecules in the voids of the ZIF-8 host. Introduction of triethylamine ($NEt_3$) during HCR, not only accelerates the reaction process resulting in a high amount of Perylene@ZIF-8 product (see Figure S1 in SI), but also it provides additional control over the morphology of the product, specifically the downsizing of the 3-D framework architecture of the ZIF-8 material to derive two-dimensional (2-D) nanoplates (Figure 1).

Figure 1 shows the 2-D morphologies of the Perylene@ZIF-8 compound we obtained, where transmission electron microscopy (TEM) and atomic force microscopy (AFM) were applied to characterize the microstructural features of the nanoplates. The lateral size of the nanoplates of Perylene@ZIF-8 varies from ~15 nm to ~150 nm, however, their thickness ranges from ~6 nm to ~20 nm (Figure 1a, b). The TEM micrographs (Figure 1c, d) revealed that, in fact the HCR product consists of an aggregation of intertwining finer scale nanoplates; see additional micrographs in Figure S2-S4 (in the SI). Notably, the aggregated nanoplates will readily transform into a monolith simply by drying at 90°C under vacuum for 6 hours (Figure S1), which can be attributed to the densification of the intertwined 2-D nanoplates.

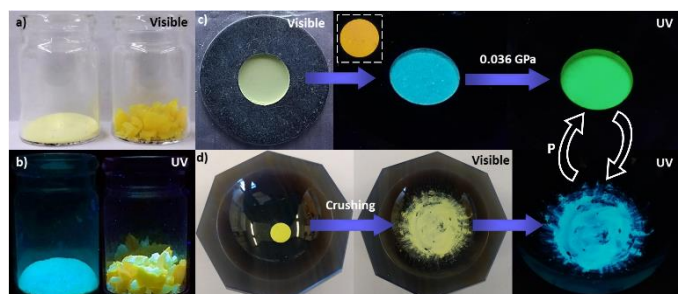

**Fig. 2** (a) Fine powder and monolithic form of Perylene@ZIF-8 seen in the visible light, and (b) their distinctively different emissions observed under the 365 nm UV excitation. (c) Reversible mechanoluminescent behavior (442 ⇌ 502 nm) of Perylene@ZIF-8: a pale-yellow powder (hand pressed into a metallic mold Ø13 mm) seen under visible light and under 365 nm UV excitation, a red-shift in emission to green 502 nm can be seen upon application of 0.036 GPa pressure. Inset shows the deep-orange emission of a pristine Perylene pellet under UV, for comparison. (d) A dark-yellow pellet observed under visible light, showing the recovery of the light-blue 442 nm emission (under UV) after being crushed to recover a fine powder form.

Interestingly, we discovered that the optical properties of the monolithic form of Perylene@ZIF-8 and its fine powder counterpart (*viz*. ground monolith), are distinctively different as summarised in Figure 2. It can be seen that the monolith material emits a light yellow-green emission, when excited under a 365 nm UV light (Figure 2a, b). But grinding of monolith into a fine powder resulted in a blue shift to emit in the lower wavelength. This drastic change in emission behavior due to mechanical grinding effect has inspired us to investigate the phenomenon in detail. We found that, as depicted in Figure 2c, the fine powders experienced a red-shift (442 nm→502 nm)

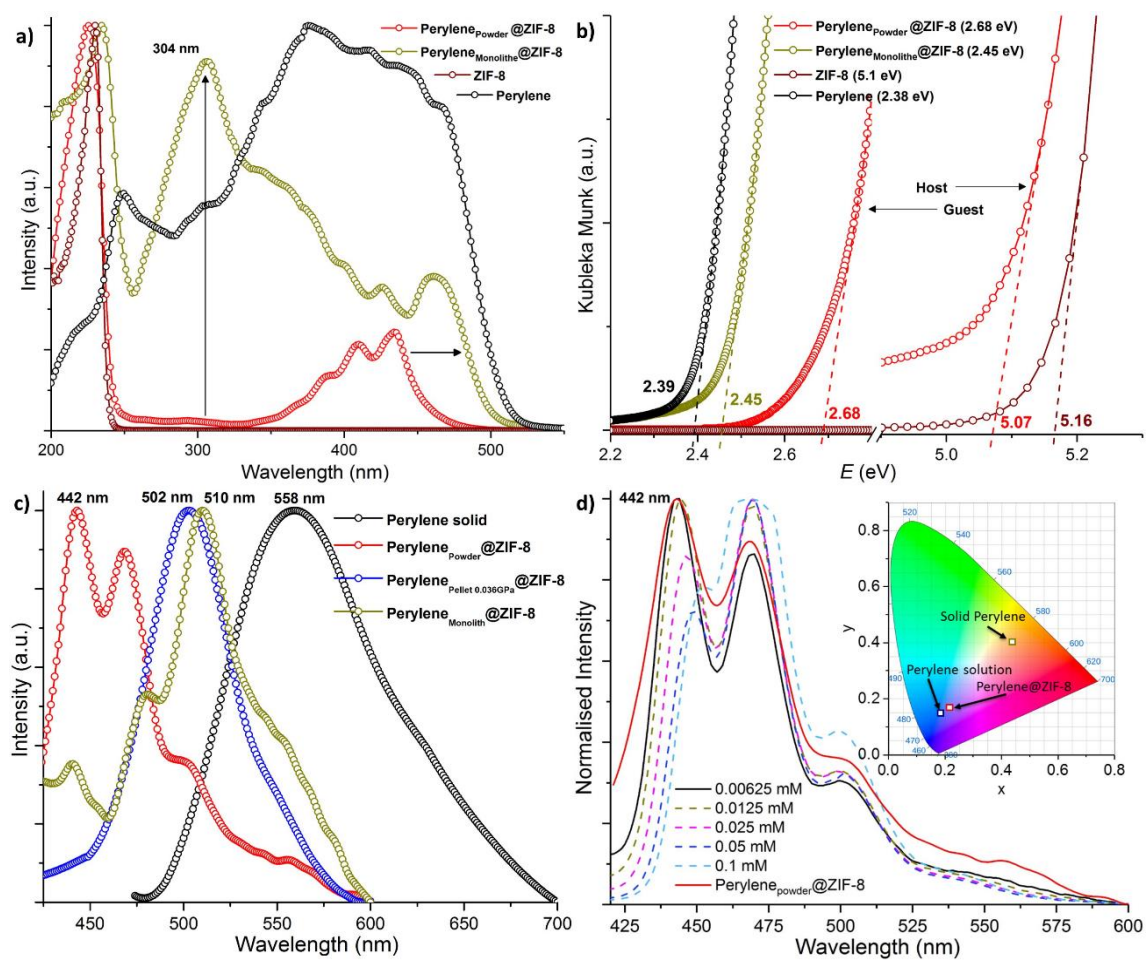

**Fig. 3** (a) Comparison of absorption behaviour obtained from diffuse reflectance of the pristine host (ZIF-8), pristine guest (solid Perylene), and host-guest system (Perylene@ZIF-8). (b) Kubelka-Munk (KM) function for determining the band gaps based on the photon energy intercepts. (c) Fluorescence emission spectra of the pristine Perylene compared with Perylene@ZIF-8 compounds when excited at 365 nm. (d) Emission spectra of different concentrations of Perylene solution (in dichloromethane) and solid-state Perylene$_{powder}$@ZIF-8. Inset: CIE 1931 colour chromaticity diagram showing the drastic effect of Perylene confinement within ZIF-8 in solid state, and how this behaviour is resembling the dilution of Perylene solution at a very low concentration of 0.00625 mM.

when being mechanically compressed into a pellet utilizing a relatively low pressure of less than ~40 MPa. Conversely, we show that by breaking the pellet down into a fine powder produced a blue-shift (502 nm→442 nm) instead, as illustrated in Figure 2d. This effect is reversible upon the application and removal of mechanical forces. In the sections below, we shall interrogate the mechanoluminescent mechanism underpinning Perylene@ZIF-8, using UV-Vis spectroscopy, theoretical band gap calculations, and X-ray diffraction techniques.

## 2. Photophysical Characterisation to Reveal Host-Guest Interactions and Energy Transfer

We performed diffuse reflectance spectroscopy to correlate the mechanoluminescence behavior to salient changes in the molecular structure of the host-guest assembly. Figure 3a shows a broad absorption peak detected between 255-510 nm for the Perylene@ZIF-8 monolith, which is completely absent in the pristine ZIF-8; this result confirms the successful confinement of Perylene (guest) within the spherical voids of ZIF-8 (host). Importantly, the spectra show significant changes observed in the absorption properties of Perylene@ZIF-8, when it was converted from a monolith into powder form (by grinding in mortar). More precisely, the distinct absorption band at 304 nm has experienced a sharp decline in its intensity when the monolith was crushed into a powder form. This finding indicates that the monolith has a pronounced π−π* transition because of the densely-packed arrangement of the nanoplate aggregates, which disappears after being converted into a loosely-packed powder form. This notion is further supported by: (i) the disappearance of the broad absorption band associated with the Perylene guest species, which transformed into multiple more well-defined peaks between 380-450 nm when the monolith was crushed into powder. (ii) The absorption peak observed at 225 nm corresponding to the pristine ZIF-8 structure has blue-shifted by ~10 nm when converting the monolith into powders, again signifying a decline in its nominal packing density.

Unlike the absorption spectra of pure Perylene in the solid-state, we note that spatially confined Perylene has absorption bands akin to the ones observed in the liquid solution.[49] Allendorf et al.[50] described the "solution-like" optical properties in the case of organic linkers in solid-state as a result of isolation in space due to MOF formation. Similar effects are observed in the current host-guest system, but the difference here is that the guest molecules (not linkers) have been isolated in space through 3-D partitioning afforded by the MOF porosity. Weak intermolecular interactions, such as π–π interactions could modify the optical behavior. Being structurally planar and aromatic in nature, the Perylene molecule may readily establish π–π stacking, therefore a deep-orange emission is observed for the solid Perylene powder (Figure 2c inset). Similar effects have been reported in organic molecules due to formation of the J (or H) type of aggregates.[51] Importantly, the ZIF-8 host environment offers tuning of the weak interactions of the confined Perylene guest emitters. In our previous report, we mentioned the ability of ZIF-8 for making weak interactions via the 2-methylimidazole (mIm) linkers.[23] Likewise, in this case the confined Perylene molecules could establish CH–π and/or π–π interactions with the adjacent imidazole rings. In fact, the resultant broad emission band of the monolith from 430 nm to 600 nm (Figure 3c) can be explained by the presence of weak host-guest interactions elucidated above.

Direct evidence of intermolecular weak interactions was witnessed from modifications to the band gap values, determined from the diffuse reflectance data. Figure 3b shows how the band gap increases from the monolith (2.45 eV) to a fine powder (2.68 eV). Electronic properties of the host framework (ZIF-8) was found to be affected by guest confinement, with its band gap decreasing from 5.16 eV to 5.07 eV. Conversely, the band gap of the Perylene guest molecule has increased due to spatial confinement, compared to that of a pure Perylene solid.

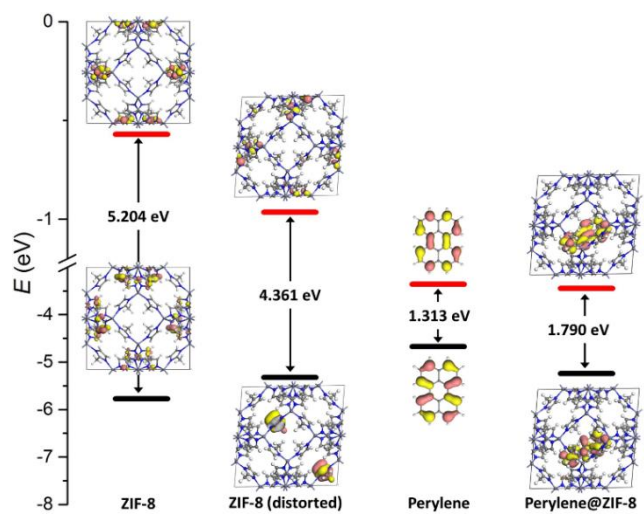

**Fig. 4** Effect of host-guest confinement on the band gap of Perylene was calculated theoretically, which followed the trends observed experimentally in Figure 3. Symbols used: HOMO (black) – LUMO (red) for the energy levels; dark pink and yellow iso-surfaces representing the positive and negative charges of the molecular orbitals.

Figure 4 presents the theoretical band gap calculations confirming this outcome, where the calculated band gap of Perylene was found to rise from 1.313 eV to 1.793 eV when it is positioned in a confined pore environment. However, the calculated band gap of the distorted ZIF-8 host (see §3 for details about the distorted ZIF-8 structure) was found to be relatively lower (4.361 eV) compared with that of a cubic ZIF-8 structure (5.204 eV). A good qualitative agreement in terms of band gap values can be seen, between the experiments and theoretical trends associated with host-guest confinement effects. These new results illustrate the potential of accomplishing band gap engineering exploiting the Guest@MOF confinement strategy.

From the photograph of the monoliths in Figure 2b, interestingly one may observe that there are different regions displaying light green to dark yellow emissions when subject to UV excitation. The edges of the monolith mainly emit a lighter green-blue color which can be assigned to the emission of 'loosely packed' nanoplates of Perylene@ZIF-8; while a darker yellow-green emission can be assigned to regions containing 'densely packed' nanoplates. These varying color characteristics are detectable in the emission spectrum of the monolith shown in Figure 3c, corresponding to the spectral hump and shoulder located at 440 nm and 479 nm, respectively. We reasoned that additional small humps observed at, for instance 543, 554, 568 and 581 nm, could arise due to entrapped Perylene molecules between the aggregation of 2-D nanoplates (but not within ZIF-8 pores). Surprisingly, the maximum emission observed at 510 nm in the case of monolithic Perylene@ZIF-8 could not be attained by the reconstituted fine powders (mechanically compressed into a pellet). Instead, an emission spectrum with only a single feature, $\lambda_{max}$ = 502 nm, was observed for the pellet of Perylene@ZIF-8 pressed under 0.036 GPa. Like absorption, fluorescence of the finely ground powder obtained from the monolithic Perylene@ZIF-8 also shows a solution-like emission behavior. Indeed, we have confirmed this phenomenon in Figure 3d, in which the emission spectra of different concentrations of Perylene solutions in dichloromethane revealed that the aggregation of molecules in a solution state causes merging of peaks at around 450 nm, but upon dilution the merged peaks became more and more separated to eventually give a maximum emission at 442 nm. This is in good agreement with the solid-state spectra of Perylene$_{powder}$@ZIF-8. Particularly, the heavily diluted Perylene with a concentration of 0.00625 mM shows less influence from its neighboring molecules, which is precisely the "solution-like" effect observed in the solid-state when the Perylene molecules have been spatially separated through nanoconfinement within the ZIF-8 host framework. Figure 3d clearly shows the transition of emission behavior of concentrated and diluted solutions and how it finally overlaps (at 442 nm) with the solid-state emission spectrum of Perylene$_{powder}$@ZIF-8.

In Figure 3c, emission of the reconstituted pellet at 502 nm suggests that the dense packing achieved in the original monolith (obtained from drying a wet sample post synthesis), cannot be fully recovered upon strain relaxation after being ground into a powder form. Therefore, the emission of monolith observed at a relatively higher wavelength of 510 nm, can be attributed to the greater charge transfer between densely aggregated nanoplates. We propose a mechanism behind the mechanoluminescence of Perylene@ZIF-8 nanoplates, as illustrated in Figure 5. It is reasonable to expect some microscopic sized voids to be present between the randomly-oriented nanoplates in powder form (Figure 5a). After application of an external pressure during the pelletisation process its packing will improve, offering better inter-nanoplate interaction to enhance electronic charge delocalisation between the adjacent nanoplates (Figure 5b), facilitated by the electron-rich Perylene guest entrapped in the pores of the sodalite cages of ZIF-8 host.

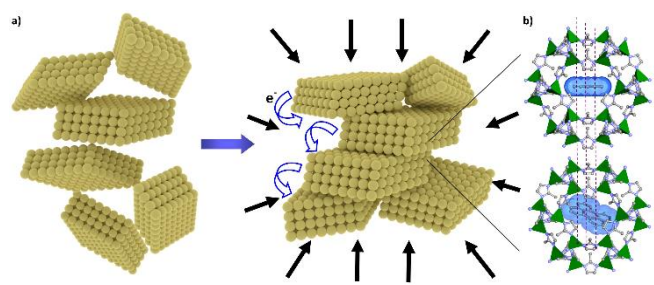

**Fig. 5** Schematic representing (a) the randomly oriented nanoplates of Perylene@ZIF-8, with air gaps present in the powder form due to its loosely-packed configuration. Upon compaction by an externally applied hydrostatic pressure, the nanoplates rearrange themselves to become more densely aggregated together. (b) Proposed energy transfer pathway of two adjacent nanoplates attributed to the "molecular wire effect", facilitated by the π–π interactions connecting the mIm…Perylene…mIm molecular pathways.

The modification of the electronic structure of the host-guest assembly is presented in Figure 4, calculated using density functional theory (DFT). Theoretical band gap energy levels give additional insights into the plausible energy transfer pathway, from the LUMO (lowest unoccupied molecular orbital) of ZIF-8 to the LUMO of Perylene upon photoexcitation. The HOMO (highest unoccupied molecular orbital) hybrid orbitals of Perylene@ZIF-8 show an electronic distribution that is highly localised on the Perylene molecules, with limited sharing with the C=C bonds in the imidazole moiety of the ZIF-8 host. The LUMO of Perylene@ZIF-8 is exclusively occupied by the Perylene molecule, indicating the propensity for energy transfer between the hybrid orbitals of the host-guest assembly and the guest species confined within. The role of the flexible nature of ZIF-8 framework can be seen in the reduction of band gap energy (~1 eV) upon shear distortion of the host structure,[47] which is again important in the context of host-guest assembly for establishing intermolecular interactions upon application of external pressure. It is well documented in literature that planarity of molecules will contribute to the reduction of band gaps by facilitating conjugation of adjacent molecules.[52] On this basis we suggest that, under pressure or mechanical stress the elasticity of ZIF-8[47] permits better planarity to be established between the local intermolecular alignment of mIm…Perylene, yielding the "molecular wire effect"[53] (Figure 5b) that could account for the red shifts we observed in the absorption and emission spectra (Figure 3).

**3. Structural Evolution as a Function of Compressive Stress**
X-ray diffraction (XRD) technique was used to characterize structural changes in the ZIF-8 host. It can be seen in Figure 6a that, the XRD pattern of Perylene@ZIF-8 monolith showed its crystalline nature, but with the presence of relatively broad peaks. Upon grinding of the monolith for ~5 minutes (*via* mortar and pestle) to yield a 'powder' sample, we detected sharper Bragg peaks that are coinciding with the simulated pattern of an ideal ZIF-8 structure. However, in addition to all the matching primary peaks, the splitting and broadening of peaks were observed suggesting the reduction of ZIF-8's cubic cell symmetry. Hence, we performed Pawley structure refinement on the experimental powder data, to establish the precise changes to the cell parameters of the host. The results in Figure 6b and Table S2 (SI) revealed that: $a = b = c = 16.992$ Å of the pristine ZIF-8 have increased to $a = 17.339$ Å, $b = 17.127$ Å, $c = 17.072$ Å, whereas, $\alpha = \beta = \gamma = 90°$ of its cubic cell have deformed to $\alpha = 87.073°$, $\beta = 90.321°$, and $\gamma = 84.677°$. The cell expansion of the host framework can be attributed to the spatial confinement of the bulky Perylene guest molecules in the pore of ZIF-8. This is accompanied by reduction in the $\alpha$ and $\gamma$ angles, which can be explained by shear deformation[47] (inset of Figure 6b) of the 2-D nanoplate morphology from aggregation-induced mechanical strains.

Grinding (*via* mortar and pestle) of the Perylene@ZIF-8 'powder' for a further ~15-20 minutes yielded a 'fine powder' sample, exhibiting broadened Bragg peaks evidenced in Figure 6a. Here, the disappearance of peak splitting can be ascribed to the relaxation of strains in the aggregated nanoplates, while peak broadening observed is indicative of the formation of finer scale particles by crushing larger aggregates. To test the effect of grinding on emission, we have compared the emission spectra of samples ground for 5-minute (powder) vs. 20-minute (fine powder), but detected no changes. To further investigate the ability of the reconstituted Perylene@ZIF-8 fine powder to exhibit red shift beyond 502 nm, we systematically prepared a series of pellets using an increasing pressure and monitored their structural evolution by XRD (Figure 6c) and emission spectroscopy. We found that the emission of the reconstituted pellets stayed unchanged at 502 nm (Figure S6), for all pellets prepared from as low as 0.036 GPa to a maximum stress of 1.47 GPa. Results above support the notion that the aggregated nanoplates in the monolithic form are very tightly packed together (yielding emission 510 nm), which cannot be regained in the reconstituted powder pellets (less dense, thus, lower emission at 502 nm).

Detailed analysis of the relative peak intensities from XRD data of different pellets made with ascending pressures can give us a deeper insight into structural changes of the material. Figure 6c shows a decreasing trend in the intensity of the (110) plane, accompanied by a gradual increase in the intensity of the (211) plane. The ratios of the changing relative peak intensities of (110):(211) planes are summarised in Figure 6c, where the

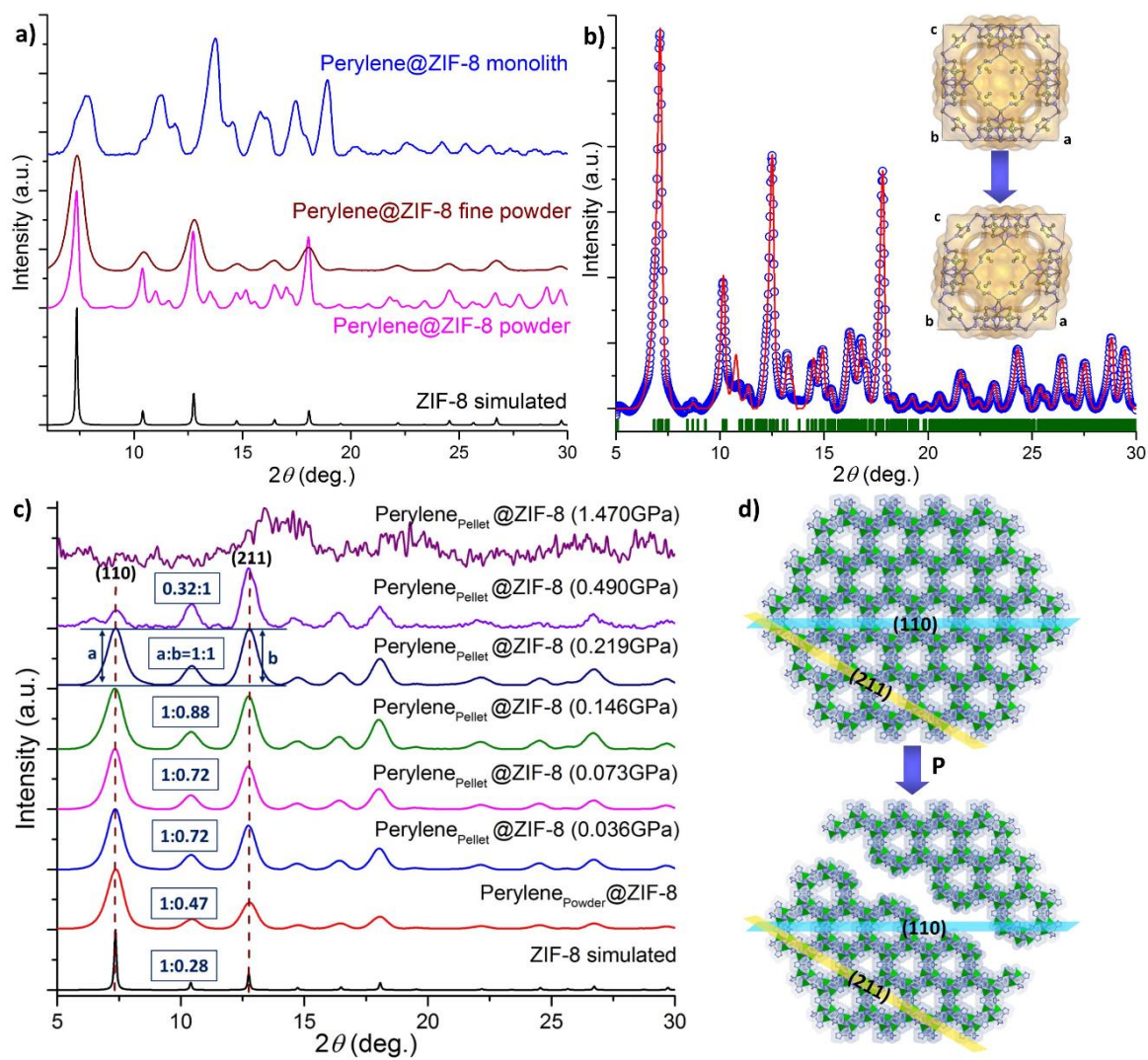

**Fig. 6** (a) Comparison of the XRD patterns of the monolithic, powder (~5 min grinding in mortar), and fine powder forms (~20 min grinding) of Perylene@ZIF-8. (b) Pawley structural refinement ($R_{wp}$ = 8.2%) of an X-ray diffraction pattern of a powder sample of Perylene@ZIF-8. Symbols used: blue circles are experimental data; red line is the calculated pattern; green bars representing the observed Bragg reflections. The inset shows the structural distortions of ZIF-8 viewed along the *b*-axis. (c) Systematic XRD study to monitor changes in relative peak intensities of the {110}- and {211}-oriented crystallographic planes upon compression of the fine powdered Perylene@ZIF-8 sample using different pressures. (d) Proposed structural fracture mechanism endorsed by XRD data, suggesting the cleavage of Perylene@ZIF-8 nanoplates across the {211} planes due to shearing deformation attributed to compressive stress, eventually causing amorphisation of the compound at ~1.5 GPa.

initial ratio rises from 1:0.47 for a ground fine powder sample to attain 1:1 for the pellet pressurised at 0.219 GPa. Then by more than doubling the pelleting pressure to 0.49 GPa, we obtained 0.32:1 alongside the appearance of extra diffraction peaks. Eventually at a significantly higher pressure of 1.47 GPa, this led to the disappearance of all the Bragg peaks associated with ZIF-8 because of framework amorphisation.[54] Figure 6d illustrates the proposed fracture mode of ZIF-8 that can account for such a phenomenon, in which the cleavage planes oriented along the {211} facets rupturing by shear stresses. In fact, this failure mechanism is reminiscent to that reported for ZIF-8 subject to a high-rate impact mechanical deformation, which also led to formation of the preferred {211}-oriented planes.[55] Significantly, all the pellets described above showed reversible mechanoluminescence behavior. Surprisingly, the amorphised sample prepared at 1.47 GPa remains mechanoluminescent by retaining its emission at 502 nm; this finding is suggesting that the spatial confinement of Perylene within the pores of ZIF-8 remains effective despite the host framework losing its long-range periodicity *via* amorphisation. Finally, thermogravimetric analysis (TGA) of the Perylene@ZIF-8 powder (Figure S7) indicated that the thermal stability of the monolithic form is higher than its powder form by ~200°C. This finding is in support of denser arrangement of nanoplates in monolithic form possessing a greater thermal stability than its fine powder form.

## Conclusions

In summary, we have shed new light on the use of orderly-arranged voids conferred by MOF compounds acting as 3-D

nanoscale scaffolding, to spatially isolate emissive guest molecules for accomplishing a unique "solution-like" luminescent behavior. In our materials design, we have employed an optically-inactive MOF host (sodalite cage of ZIF-8) for the spatial confinement of a bulky polycyclic fluorophore guest (Perylene), resulting in a novel Guest@MOF system — Perylene@ZIF-8 — with switchable emission properties (442 ⇌ 502 nm) when subject to a reversible mechanical stress or hydrostatic pressure. We have harnessed the high-concentration reaction (HCR) synthetic strategy for morphology control, yielding nanoplates of an inherently 3-D ZIF-8 structure co-assembled with the Perylene guests. Remarkably, such nanoscale 2-D configuration permits facile aggregation and separation of Perylene@ZIF-8 nanoplates under stress (~10s MPa), which is responsible for the reversible mechanoluminescence phenomenon we recorded. Furthermore, we discovered that the structural transition from a crystalline to amorphous phase could still retain its mechanoluminescence properties, thereby opening up an exciting route for fabricating functional amorphous materials constituting photonic Guest@MOF nanocomposites. Together, our new findings and innovative approach towards mechanoluminescence will be valuable in expanding the frontier of MOF science, especially in the pursuit of next-generation host-guest nanomaterials exhibiting tunable optoelectronic and sensing properties.

## Acknowledgments

This research was funded by the Samsung Advanced Institute of Technology (SAIT) GRO, and the Engineering and Physical Sciences Research Council, EPSRC RCUK (EP/N014960/1 and EP/K031503/1). We are very grateful to Dr. Gavin Stenning and Dr. Marek Jura at the R53 Materials Characterisation Lab in the ISIS Rutherford Appleton Laboratory, for the provision of X-ray diffraction facilities. We thank the Research Complex at Harwell (RCaH), Oxfordshire, for allowing access to the TEM and UV-Vis facilities.